# The retrieval effectiveness of search engines on navigational queries

Dirk Lewandowski
Hamburg University of Applied Sciences – Faculty Design, Media and Information, Department Information, Berliner Tor 5, D – 20249 Hamburg, Germany
E-Mail: dirk.lewandowski@haw-hamburg.de

**Abstract**

Purpose – To test major Web search engines on their performance on navigational queries, i.e. searches for homepages.

Design/methodology/approach – 100 real user queries are posed to six search engines (Google, Yahoo, MSN, Ask, Seekport, and Exalead). Users described the desired pages, and the results position of these is recorded. Measured success N and mean reciprocal rank are calculated.

Findings – Performance of the major search engines Google, Yahoo, and MSN is best, with around 90 percent of queries answered correctly. Ask and Exalead perform worse but receive good scores as well.

Research limitations/implications – All queries were in German, and the German-language interfaces of the search engines were used. Therefore, the results are only valid for German queries.

Practical implications – When designing a search engine to compete with the major search engines, care should be taken on the performance on navigational queries. Users can be influenced easily in their quality ratings of search engines based on this performance.

Originality/value – This study systematically compares the major search engines on navigational queries and compares the findings with studies on the retrieval effectiveness of the engines on informational queries.

Paper type – research paper

**Introduction**

The importance of search engines for finding relevant information on the World Wide Web is indisputable. Consequently, there is a wide interest among researchers in the quality of search engine results. However, most studies operate under the classic assumption of a "dedicated searcher" who is willing to go through vast results lists to pick all relevant results and evaluate them. While this only holds partly true for Web searches (where the users are, in most cases, satisfied with just a few results), we find that in Web searching, users pose queries for another purpose, too—to navigate to a certain Web page they already know or assume to exist. We believe that the success for this kind of queries is a critical success factor for Web search engines.

In evaluation of the retrieval effectiveness for informational queries, we find that some groups of search engines exist, where Google and Yahoo are clearly leading the field (Lewandowski, 2008). The question is whether these results also hold true for navigational queries. It could be that one search engine is superior for these kinds of queries. One question in our overall search engine quality measurement framework (Lewandowski & Höchstötter, 2008) is why users perceive Google as the best search engine, as shown in market figures (e.g., "Suchmaschinen-Marktanteile," 2008; "Webhits," 2007) and stated in user surveys (e.g., Machill, Neuberger, Schweiger, & Wirth, 2003; Schmidt-Maenz & Bomhardt, 2005). This assumption could not be proved solely in terms of results quality. Explanations (apart from design issues and convenience) could be as follows:
1. Google indeed delivers the best possible results. However, our comparisons on informational queries show that Google only delivers better results than its next best competitor for the first few results positions (and only if the combination of relevant results descriptions and relevant results is taken into account). When more results positions are considered, the Yahoo's results are even with or better than Google's (Lewandowski, 2008).
2. Google delivers the best results descriptions. As users are not able to compare systematically the results of different search engines, they must rely on what they see. And, as they usually only consider one or a few results per query (Granka, Joachims, & Gay, 2004; Keane, O'Brien, & Smyth, 2008), they must rely on the impression they get from the results lists. In our study on the interplay of results descriptions and actual results (Lewandowski, 2008), we were able to show that Google delivers the highest ratio of relevant results descriptions, while the results themselves are clearly inferior to the descriptions.
3. Google performs best on navigational queries. With such queries, it is easy for a user to determine whether a given result set is good. If the desired Web site is shown at the top position (or at least at one of the first positions), the information need is satisfied. If the user has to scroll down to find the desired page—or worse, the page does not show up at all—the user is disappointed and may not use the search engine again.

The present study deals with the last assumption. Our research is guided by the question of whether Google actually performs best on navigational queries, which would be one explanation for why users prefer Google. Users are able to judge the performance of different search engines on navigational queries, while they are restricted in judging the quality of a *results set* on informational queries.

This paper is organised as follows. First, we further explain the concept of navigational queries and give a concise literature review. Then, we present our research questions. In the next section, we describe our methods for data collection. After that, results are presented and discussed. The paper closes with a conclusion and suggestions for further research.

**Literature review**

Many tests have been conducted on the retrieval effectiveness of Web search engines. Newer studies are Griesbaum (2004), Lewandowski (2008), and Véronis (2006); for an overview, see Lewandowski (2008). However, in most cases, these studies use informational queries, that is, queries where the user wants to find at least some documents on his search topic. The information need is usually not satisfied with just one result, and the user does not know in advance which document (or, in the Web context, page) will provide the most useful result.

In contrast, the concept of navigational queries is based on known-item searches in library collections (Kilgour, 1999) or, more generally, on the differentiation between problem-oriented information needs and concrete information needs (Frants, Shapiro, & Voiskunskii, 1997, p. 38). The user knows about an item in a collection and wants to retrieve it. Therefore, only one result would be relevant. The search is completed when the desired item is found.

Adapting this concept to Web queries and contrasting it with the usual query type, Andrei Broder (2002) presents a taxonomy of Web search that focuses on the goals a user pursues with a query. The author distinguishes between navigational, informational, and transactional queries.

With informational queries, users want to find information on a certain topic. Such queries usually lead to a set of results rather than to just one suitable document. Informational queries are similar to queries sent to traditional text-based information retrieval systems. According to Broder, such queries always target static Web pages. The term "static" here should not refer to the technical delivery of the pages (e.g., dynamically generated pages by server side scripts like php or asp), but rather to the fact that once the page is delivered, no further interaction is needed to get the desired information.

Navigational queries are used to find a certain Web page the user already knows about or assumes exists. A different expression for such queries is "homepage finding" (as used in TREC). As Hawking & Craswell (2005, p. 219) put it, "The judging criterion is, 'Is this the page I wanted?—that is, the home page of the entity I was thinking of or the page I named?"

Typical queries in this category are searches for a homepage of a person or organization. A typical example is the search for a company (e.g., "Daimler Chrysler"). Navigational queries are usually answered by just one result; the information need is satisfied as soon as this one right result is found. However, not all queries for people are navigational. Rose and Levinson (2004) are of the opinion that most queries for people are, in fact, not. They point out that "a search for celebrities such as Cameron Diaz or Ben Affleck typically results in a variety of fan sites, media sites, and so on: it's unlikely that a user entering the celebrity name as a query had the goal of visiting a specific site." This may be true, but these queries cannot be seen as informational because one cannot assume that the user wants to read a variety of documents. Therefore, we decided to classify queries for celebrities as navigational.

The results of transactional queries are Web sites in which a further interaction is necessary. A transaction can be the download of a program or file, the purchase of a product, or a further search in a database.

Based on a log file analysis and a user survey (both from the AltaVista search engine), Broder finds that each query type stands for a significant number of all searches. Navigational queries account for 20–24.5 percent of all queries, informational queries for 39–48 percent, and transactional queries for 22–36 percent.

The results from Rose and Levinson (2004) are only in part comparable to Broder's original study, mainly because of different definitions of navigational and transactional queries. According to Rose and Levinson, informational queries account for 61 to 63 percent of all queries, transactional queries for 21 to 27 percent, and navigational queries for only 11 to 15 percent.

The division of the queries into three types should be seen as seminal, because it was the first attempt to differentiate between user intents expressed in queries. Further work investigated the possibility of assigning an intention to a query automatically (Kang & Kim, 2003).

In a study using data from three German search engines, Lewandowski (2006) finds that navigational queries account for around 40 percent of all queries. Jansen, Booth, & Spink (2008) classify queries automatically. They find a quite lower ratio of navigational queries in their data sets (10.2 percent). However, it is unclear whether this result depends on the data set, changing user behaviour, or their approach to automatic classification.

While the exact numbers on the ratio of navigational queries differ from one study to another (see Table 1), the clear result of all studies is that navigational queries account for a noteworthy number of queries.

Table 1: Distribution of query types in the different search engine studies

| Data set | Informational | Navigational | Transactional |
|---|---|---|---|
| AltaVista 2002 (Broder, 2002) | 39−48% | 20−24.5% | 22−36% |
| AltaVista 2004 (Rose & Levinson, 2004) | 61−63% | 11−15% | 21−27% |
| German search engines 2005 (Lewandowski, | 45% | 40% | 15% |
| 2006) | | | |
| Several data sets (Jansen et al., 2008) | 80.6% | 10.2% | 9.2% |

* automatic classification

Several retrieval measures were developed to calculate the retrieval effectiveness for navigational queries; see MacFarlane (2007, p. 354).

Success *N* asks whether the right result appears within the first N results (Craswell & Hawking, 2005). In the Web Track of TREC, success is calculated at 1, 5, and 10. Mean reciprocal rank is a standard measure in TREC. It uses a descending scale, that is, the higher the position in which the result is found, the higher the score. Percentage top *N* calculates the proportion of queries where the right answer was found in the top *N* hits, while percentage fail *N* calculates the proportion of queries where no right answers were found in the top *N* hits (MacFarlane, 2007).

**Research questions**

We defined some research questions guiding our presentation and discussion of results, as follows:

RQ1: Which search engine is able to produce the most relevant results on the first position?

RQ2: How many queries remain unanswered with a certain search engine?

RQ3: Do the search engines face problems putting the relevant result on the first position, that is, to what degree are relevant results found on the lower ranks of the results list?

**Methods**

A group of students was introduced to the concept of navigational queries and were asked for their last such query posed to a search engine. We collected 100 German-language queries from individual students, each with the exact query, a short description of the information need, and (when remembered) the URL of the desired page.

For our comparison, we selected six search engines based on the restriction that they should provide their own index and that they are of significance at least for a certain language area. We chose Google, Yahoo, MSN/Live.com, Ask, Exalead, and German search engine Seekport. With the exception of Exalead, these are the same search engines we used in our other tests. Data was collected in January 2008.

Each query was posed to the German-language interface of each search engine under investigation and the position of the correct results was recorded. If the correct page could not be found within the first 10 results, the search was abandoned.

In contrast to investigations on informational queries, relevance judgements were not a problem in our study. It is obvious which result is relevant and which results are not. Therefore, we had no problem in selecting jurors, making results anonymous, or changing the sequence of the results. One person gave all the relevance judgements.

Some large companies provide regional or language-specific homepages that are on the same server, instead of separate servers (top-level domains), for each country. For example, the German homepage of Apple Computers is www.apple.com/de instead of www.apple.de. Just "www.apple.com" would not be the correct result in our test, as our queries were German and in case of multiple language versions, we assumed that the German one would be desired.

Due to this situation, we define a desired page in our investigation as a page that either matches the exact URL as stated by the user or as a page that matches all criteria given by the user in the description of his or her information need.

We calculate different retrieval measures for our data, as described in Craswell & Hawking (2005), for navigational queries: mean reciprocal rank (using a scale from 1.0 for the first position, 0.5 for the second, etc. Results on position 6 or lower receive a score of 0), and success $N$. In addition, we measure the gain ratio, that is, the ratio a search engine improves from only the first result considered to all results investigated considered.

**Results**

*Number of queries answered/unanswered*

First, it is interesting to see how many queries the search engines are able to answer satisfactorily at all, that is, regardless of the results position. Fig. 1 shows how many queries remain unanswered with the individual search engines. While the three major search engines (Google, Yahoo, and MSN) each fail in around 10 percent of the queries; results are much worse for the smaller engines Ask, Seekport, and Exalead. This may have to do with optimisation of the engines for certain languages. However, it is interesting to see that Seekport—the only genuine German search engine of any importance—performs worst in our test. Fifty-seven percent of all queries remain unanswered with this engine.

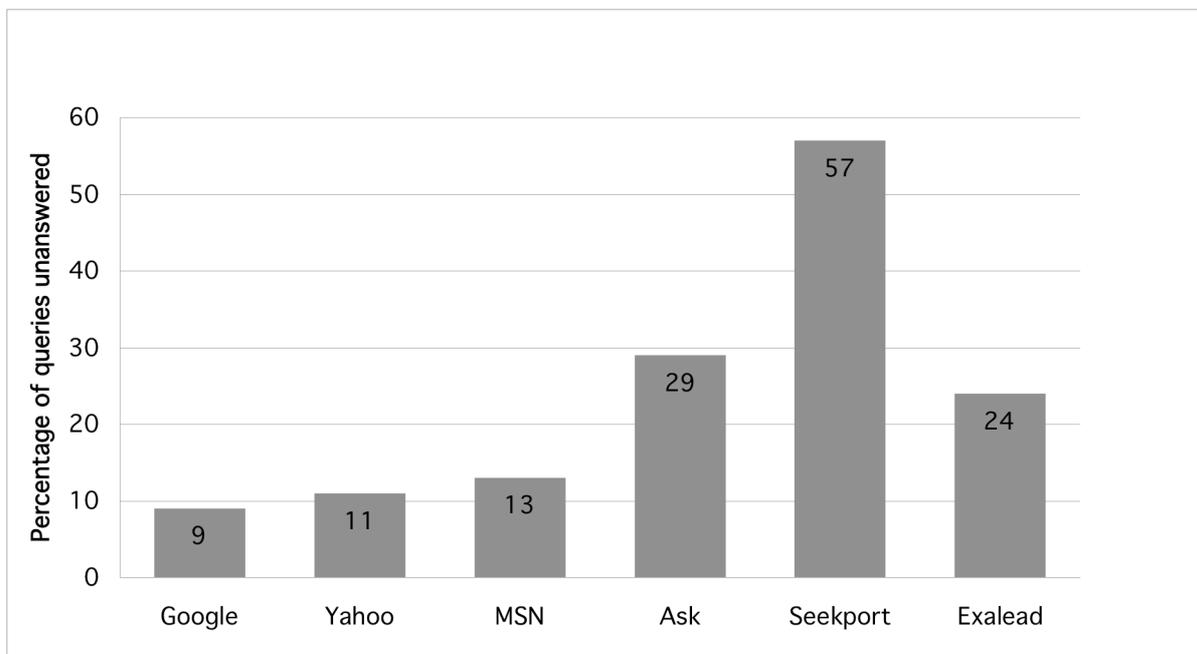

Fig. 1: Number of queries unanswered by the individual search engines

*Unanswerable queries*

Looking at the best three search engines, with around 10 percent failure rates, it is interesting to know whether these queries are simply not answerable. For the purpose of our study, we define unanswerable as not answered by any of the search engines under investigation. The results show that there are only four queries that none of the engines was unable to answer. However, on the other side, only one-third of all queries were answered by all of the engines. Details for all engines can be found in Table 2.

Table 2: Number of queries answered by n search engines

| Found by (number of engines) | none | 1 | 2 | 3 | 4 | 5 | all |
|---|---|---|---|---|---|---|---|
| Number of queries | 4 | 1 | 2 | 10 | 0 | 20 | 33 |

*Success N*

The next question is whether the engines are able to put the desired page on the first rank of the results page and in how many more cases the query is answered with the result to be found on a lower rank of the top 10 results list. For example, Google finds 92 of the desired pages in total, but only puts 84 of them on the first rank. The gain ration from the first result to the complete set of ten results is under 10 percent. This means that the effort for a searcher to scan all ten results on the list is relatively high for the relatively low gain.

As expected, the gain ratios for the engines with a lower ratio of the desired result on the first rank are higher. It can be seen from the data that, while MSN performs nearly as well as Google and Yahoo when all ten results positions are considered, this engine has a much lower ratio of queries when the desired page is in the first results position. While Google and Yahoo put the desired results in the first position in 84 and 82 percent, respectively, MSN only succeeded in only 75 percent of the cases. Considering the importance of putting the desired page on the first position (see above), this difference is highly significant. It shows that the problem with this search engine is not to find the desired page at all, but to rank it according to the user's intention.

Table 3: Success *N* and gain ratio (form first result to complete results set)

| Search engine | Google | Yahoo | MSN | Ask | Seekport | Exalead |
|---|---|---|---|---|---|---|
| Success 1 | 84 | 82 | 75 | 60 | 36 | 66 |
| Success 2 | 89 | 84 | 84 | 64 | 39 | 72 |
| Success 3 | 89 | 84 | 85 | 65 | 41 | 76 |
| Success 4 | 89 | 86 | 86 | 67 | 42 | 76 |
| Success 5 | 90 | 88 | 86 | 68 | 42 | 76 |
| Success 6 | 90 | 88 | 88 | 70 | 43 | 76 |
| Success 7 | 90 | 88 | 88 | 71 | 43 | 76 |
| Success 8 | 91 | 88 | 88 | 72 | 44 | 77 |
| Success 9 | 92 | 88 | 88 | 72 | 44 | 77 |
| Success 10 | 92 | 90 | 88 | 72 | 44 | 77 |
| Gain ratio from $1^{st}$ to $10^{th}$ | 9.52% | 9.76% | 17.33% | 20.00% | 22.22% | 16.67% |

In Fig. 2, the gains from the first result to the tenth result are plotted for each engine. As expected, the graphs are not steep. With relatively high values, even for the first results with most engines, there is not so much to gain. Considering only the top three engines (Google, Yahoo, and MSN), we find that only MSN improves a noteworthy amount when the whole results set is considered.

There are some interesting differences with the top search engines. When considering the first two or three results, one finds that Google performs better than its competitors. When considering four or more results, the differences are not significant anymore. It seems that the perception of Google as the best search engine for navigational queries depends on the number of results a user considers. This amount is heavily dependent on the screen/window size, that is, the number of results a user sees without scrolling down (Granka et al., 2004).

With Seekport, the gain ratio (from a low staring point) is relatively high. However, this engine in general has problems in bringing the desired page into the top 10 results. Whether this has to do with the ranking or whether the desired pages are simply not in this search engine's index cannot be answered from our data.

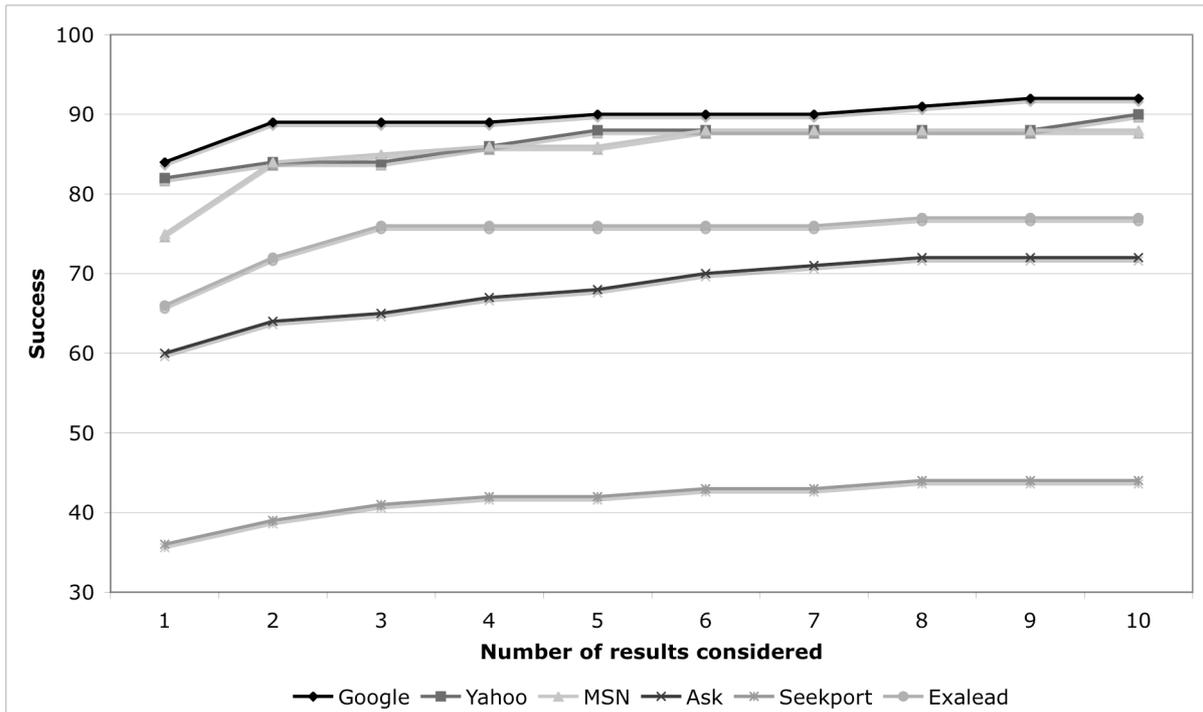

Fig. 2: Gain from first result to complete results set

*Mean reciprocal rank*

As stated in the methods section, we also calculated the mean reciprocal rank. Scores bisect from one results position to the next, and up to five results positions are considered. Results are 0.87 for Google, 0.83 for Yahoo, 0.80 for MSN, 0.70 for Exalead, 0.63 for Ask, and 0.38 for Seekport. This confirms the results above in which success *N* was used.

**Discussion and conclusions**

The discussion follows our research questions, as stated above.
Regarding RQ1 ("Which search engine is able to produce the most relevant results on the first position?"), we find that Google and Yahoo confirm their outstanding positions as detected in retrieval effectiveness tests using informational queries (Lewandowski, 2008). As we used German queries, it is very disappointing to find genuine German search engine Seekport performing the worst in our test. However, the results show that often-criticised U.S. search engines are able to produce good results for foreign-language queries. To what extent this also holds true for other languages than German is unclear and should be investigated in further research.
It is pleasant to see that there seems to be no general problem in not finding the desired pages at all (RQ2). Only four of our 100 queries remained unanswered, that is, none of the engines was able to answer them.
Answering RQ3 ("Do the search engines face problems putting the relevant result on the first position, that is, to what degree are relevant results found on the lower ranks of the results list?"), we have to differentiate between the individual search engines. Most of the results found by Google and Yahoo are indeed listed on the first rank of the results page, and the gain ratio from only the first rank considered to all ranks considered is quite low. Both engines do improve on the first few results (Google mainly from the first to the second result, Yahoo from the first to the fifth results), which means that they are in around 90 percent of cases able to show the desired page within the results set that the user can see on the first results page without scrolling down. This is very important, as users are usually not willing to consider results not shown "below the fold" of the first screen (Granka et al., 2004) or even to consider results past the first results page of 10.

Looking at the performance of MSN, we find that, while this engine performs worse than Google and Yahoo for the first ranking position, results improve as more positions are considered. The weaker performance for the first results position could lead to the impression that this search engine is inferior to its main competitors.

However, when comparing the results for the major search engines with the results for experimental systems evaluated in TREC (Craswell & Hawking, 2005), we find that the commercial clearly outperform the experimental systems. However, it should be kept in mind that a direct comparison of these systems is not valid. Results could just give a raw comparison. Furthermore, TREC differentiates between "named pages" and "homepage results," while, in our study, both types of pages are mixed into one.

Practical implications from our research are that a user performing a navigational query should reformulate his or her query if the desired page is not to be found within the first few results. The ranking components of the commercial search engines are capable of ranking these pages within the first few results. Therefore, it is inefficient to look at further pages of the results list.

Further research is needed on the reasons why certain pages are found by just one or a few search engines, but not by all. A qualitative analysis could help find the reasons why some pages are more difficult to rank for our queries than others. As stated above, our results are just valid for German-language queries. Further research should focus on other languages, too.

**References**

Broder, A. (2002), "A taxonomy of Web search", *SIGIR Forum,* Vol. 36 No. 2, pp. 3-10.

Craswell, N., and Hawking, D. (2005), "Overview of the TREC-2004 Web Track" available at: trec.nist.gov/pubs/trec12/papers/WEB.OVERVIEW.pdf (accessed 17 July 2008).

Frants, V.I., Shapiro, J., and Voiskunskii, V.G. (1997), *Automated Information Retrieval: Theory and Methods*, Academic Press, San Diego.

Granka, L.A., Joachims, T. and Gay, G. (2004), "Eye-tracking analysis of user behavior in WWW search", *Proceedings of Sheffield SIGIR*, Twenty-Seventh Annual International ACM SIGIR Conference on Research and Development in Information Retrieval, pp. 478-479.

Griesbaum, J. (2004), "Evaluation of three German search engines: Altavista.de, Google.de and Lycos.de", *Information Research*, Vol. 9 No.4.

Hawking, D. and Craswell, N. (2005), "The Very Large Collection and Web Tracks", in Voorhees, E.M. and Harman, D.K. (Eds.), *TREC Experiment and Evaluation in Information Retrieval*, MIT Press, Cambridge, Massachusetts, pp. 199-231.

Jansen, B.J., Booth, D.L., and Spink, A. (2008), "Determining the informational, navigational, and transactional intent of Web queries", *Information Processing & Management,* Vol.44 No.3, pp. 1251-1266.

Kang, I.H., and Kim, G. (2003), "Query Type Classification for Web Document Retrieval", *SIGIR Forum*, ACM Special Interest Group on Information Retrieval, SPEC. ISS., pp. 64-71.

Keane, M.T., O'Brien, M. and Smyth, B. (2008), "Are people biased in their use of search engines?" *Communications of the ACM,* Vol. 51 No. 2, pp. 49-52.

Kilgour, F.G. (1999), "Retrieval effectiveness of surname-title-word searches for known items by academic library users", *Journal of the American Society for Information Science,* Vol. 50 No. 2-3, pp. 265-270.

Lewandowski, D. (2006), "Query types and search topics of German Web search engine users", *Information Services & Use,* Vol. 26 No. 4, pp. 261-269.

Lewandowski, D. (2008), "The Retrieval Effectiveness of Web Search Engines: Considering Results Descriptions", *Journal of Documentation,* Vol. 64 [to appear].

Lewandowski, D. and Höchstötter, N. (2008), "Web Searching: A Quality Measurement Perspective". in Spink, A.; Zimmer, M. (Eds.), *Web Search: Multidisciplinary Perspectives*, Springer, Berlin, pp. 309-340.

MacFarlane, A. (2007), "Evaluation of Web search for the information practitioner", *Aslib Proceedings: New Information Perspectives,* Vol. 59 No. 4-5, pp. 352-366.

Machill, M. and Neuberger, C., Schweiger, W., & Wirth, W. (2003). "Wegweiser im Netz: Qualität und Nutzung von Suchmaschinen", in Machill, M. and Welp, C. (Eds.), *Wegweiser im Netz*, Bertelsmann Stiftung, Gütersloh, pp. 13-490

Rose, D.E. and Levinson, D. (2004), "Understanding user goals in Web search", *Thirteenth International World Wide Web Conference Proceedings*, WWW2004.

Schmidt-Maenz, N. and Bomhardt, C. (2005), "Wie suchen Onliner im Internet?", *Science Factory/Absatzwirtschaft*, 2 , pp. 5-8.

"Suchmaschinen-Marktanteile (2008)", available at: www.luna-park.de/home/internet-fakten/suchmaschinen-marktanteile.html (accessed 17 July 2008).

Véronis, J. (2006), "A comparative study of six search engines", available at: www.up.univ-mrs.fr/veronis/pdf/2006-comparative-study.pdf (accessed 17 July 2008).

"Web-Barometer", available at: www.webhits.de/deutsch/index.shtml?webstats.html (accessed 17 July 2008).